\documentclass[%
reprint,
superscriptaddress,
amsmath,amssymb,
aps,
pra,
]{revtex4-2}

\usepackage{graphicx}
\usepackage{dcolumn}
\usepackage{bm}
\usepackage{braket}
\usepackage{color}

\newcommand{\im}{\textrm{i}}  
\newcommand{\e}{\textrm{e}} 
\graphicspath{{C:/Users/h1005610.TOSHIBA/Documents/}}
\graphicspath{{images/}}
\begin{document}

\preprint{APS/123-QED}

\title{Two-qubit gate using conditional driving 
for highly detuned Kerr-nonlinear parametric oscillators
}

\author{Hiroomi Chono}
 \email{hiroomi1.chono@toshiba.co.jp}
\affiliation{%
Frontier Research Laboratory, Corporate Research \& Development Center, Toshiba Corporation, 1, Komukai Toshiba-cho, Saiwai-ku, Kawasaki-shi, 212-8582, Japan 
}%


\author{Taro Kanao}
\affiliation{%
Frontier Research Laboratory, Corporate Research \& Development Center, Toshiba Corporation, 1, Komukai Toshiba-cho, Saiwai-ku, Kawasaki-shi, 212-8582, Japan 
}%

\author{Hayato Goto}
\affiliation{%
Frontier Research Laboratory, Corporate Research \& Development Center, Toshiba Corporation, 1, Komukai Toshiba-cho, Saiwai-ku, Kawasaki-shi, 212-8582, Japan 
}%

\date{\today}

\begin{abstract}
A Kerr-nonlinear parametric oscillator (KPO) is one of the promising devices to realize qubits for universal quantum computing.
The KPO can stabilize two coherent states with opposite phases, yielding a quantum superposition called a Schr\"{o}dinger cat state.
Universal quantum computing with KPOs requires three kinds of quantum gates: $R_z, R_x$, and $R_{zz}$ gates.  
We theoretically propose a two-qubit gate $R_{zz}$ for highly detuned KPOs.
In the proposed scheme, we add a parametric drive for the first KPO. 
This leads to the $R_{zz}$ gate based on the driving of the second KPO depending on the first-KPO state, which we call ``conditional driving."
First, we perform simulations using a conventional KPO Hamiltonian derived from a superconducting-circuit model under some approximations and evaluate the gate fidelity. 
Next, we also perform numerical simulations of the two-qubit gate using the superconducting-circuit model without the approximations. 
The simulation results indicate that the conditional-driving gates can be implemented with high fidelity ($>99.9\%$) for rotation angles required for universality. 
\end{abstract}

\maketitle


\section{\label{sec:level1} 
INTRODUCTION}
Superconducting circuits (SCs) are considered as a promising platform for implementing practical quantum computation~\cite{krantz2019,kwon2021,blais2021}.
The physical properties of superconducting qubits have garnered much attention since the first realization of a superconducting qubit about two decades ago~\cite{nakamura1997,nakamura1999}. 
Inspired by cavity quantum electrodynamics (QED) in atomic physics and quantum optics, circuit QED has also been developed~\cite{wallraff2004,blais2004,arute2019,gong2021,blais2021}, and has improved technologies, such as qubit readout~\cite{wallraff2005}, parametric amplifiers and oscillators~\cite{yamamoto2008,lin2013,lin2014}, and various quantum-optics experiments~\cite{schuster2007,kono2017}. 

Moreover, a charge-based superconducting qubit robust against stray electric field noise was proposed~\cite{koch2007,schreier2008}. The qubit is called a transmon, and it has become standard as a SC qubit.  
The transmon can be regarded as a nonlinear $LC$ circuit whose nonlinearity originates from Josephson junctions, and also as an anharmonic oscillator with the Kerr effect. 

As a further development, a parametric oscillator formed by nonlinear $LC$ circuits whose Kerr nonlinearity is larger than the single-photon loss rate has been interested as a qubit, which is called a Kerr-nonlinear parametric oscillator (KPO)~\cite{goto2016,goto2016_uni,goto2019}.

Although Kerr-nonlinear oscillators with/without parametric drive were theoretically studied more than two decades ago~\cite{milburn1986,miranowicz1990,milburn1991,leonski1994,leonski1996,ralph2003}, the KPO has recently attracted intense attention again because of recent intriguing theoretical proposals for quantum annealing~\cite{goto2016,goto2019,nigg2017,puri2017q,goto2018,onodera2020,goto2020,kanao2021} and gate-based quantum computing~\cite{goto2016_uni,puri2017,goto2019,puri2017q,darwaman2021,kwon2022,kanao2021q,masuda2022,mirrahimi2014}.

A KPO can stabilize a quantum superposition of coherent states with opposite phase~\cite{goto2016,goto2016_uni,puri2017,goto2019,puri2020,darwaman2021}, which is often called a Schr\"{o}dinger cat state. 
A KPO is similar to a Josephson parametric oscillator (JPO) using a SC~\cite{lin2014}. KPOs have been implemented experimentally with superconducting quantum interference devices (SQUIDs)~\cite{wang2019,yamaji2022} or by a superconducting nonlinear asymmetric inductive element (SNAIL) transmon~\cite{grimm2020}. 
For gate-based quantum computing using KPOs, three kinds of gates are required for universality: $R_x$ gate ($X$ rotation), $R_z$ gate ($Z$ rotation), and $R_{zz}$ gate ($ZZ$ rotation)~\cite{goto2016_uni,puri2017,mirrahimi2014}. 
The $R_{zz}$ gate can be realized by controlling a linear coupling between two KPOs~\cite{goto2016_uni,puri2017}.
The single-qubit gates ($R_x,~R_z$) have been experimentally demonstrated, although the angle of one of them was limited, that is, not continuous~\cite{grimm2020}.
In contrast, the two-qubit gate ($R_{zz}$) for KPOs has not been experimentally realized yet. 

In this paper, we theoretically propose a two-qubit gate for highly detuned KPOs coupled with a fixed coupling rate.
During idle time, the coupling between the KPOs is effectively turned off because of the large detuning between the KPOs. To perform a two-qubit gate, we add a parametric drive for the first KPO, the frequency of which is set to the sum of the KPO frequencies.
This leads to the generation of photons with the second-KPO frequency through a three-wave mixing process in the first KPO, resulting in the driving of the second KPO depending on the first-KPO state, which we call ``conditional driving."
The conditional driving is physically similar to the previously demonstrated cat-quadrature readout for a KPO~\cite{grimm2020}.
The $R_{zz}$ gate can be realized using the conditional driving induced by a gate pulse of the parametric drive, which we call the conditional-driving gate.

We propose a SC model of two coupled KPOs for the $R_{zz}$ gate and derive a simple model from the SC model under some approximations.
Using these models, we show that high gate fidelity ($>99.9~\%$) can be achieved for rotation angles required for universal quantum computing. 

This paper is organized as follows. 
In Sec.~\ref{sec:sc-model}, we propose a two-qubit gate for highly detuned KPOs and show a  SC for that.
For simplicity, we start with the explanation of a single KPO and derive a conventional KPO Hamiltonian, which we call a simple model, from the SC model under some approximations.
Next, we explain a SC model of the two-qubit gate for two coupled KPOs and derive a simple model under the same approximations.
In Sec.~\ref{sec:Rzz}, using the simple and SC models, we demonstrate numerically that the $R_{zz}$ gate can be implemented with high fidelity. 
In Sec.~\ref{sec:damping}, we investigate the effect of single-photon loss for the gate fidelity.
Finally, we conclude this paper in Sec.~\ref{sec:conc}.

\section{\label{sec:sc-model}
Model}
\begin{figure}[t]
\centering
\includegraphics[width=8cm]{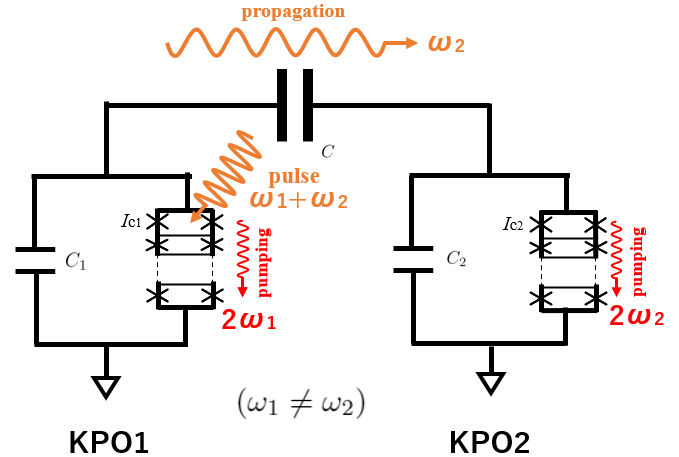}
\caption{Circuits of the two-qubit gate for KPOs using three-wave mixing (irradiated with the sum frequency). $\omega_j$ is the eigenfrequency of KPO$j~(j=1,2)$.}
\label{fig:2KPO}
\end{figure}

\subsection{
$R_{zz}$ gate and SC}
\begin{figure}[t]
\centering
\includegraphics[width=6cm]
{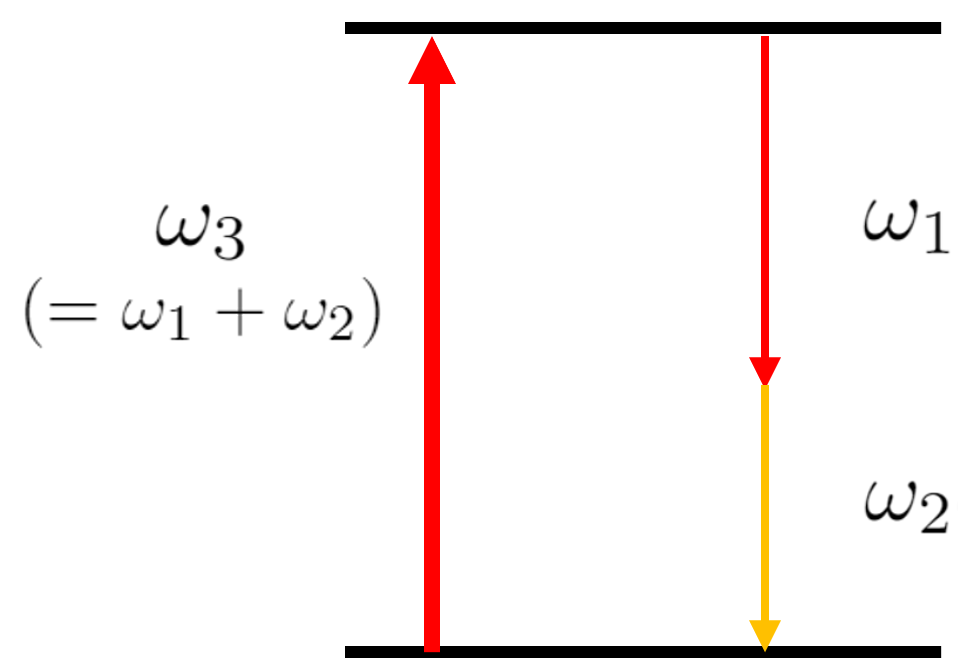}
\caption{Three-wave mixing induced by the sum-frequency $(\omega_3=\omega_1+\omega_2)$ drive in KPO1. This leads to difference-frequency generation of a photon of $\omega_2$ from a pump photon of $\omega_3$ and a KPO1 photon of $\omega_1$.}
\label{fig:sum}
\end{figure}

We propose the $R_{zz}$ gate for highly detuned KPOs.
The proposed SC is shown in Fig.~\ref{fig:2KPO}, where two KPOs are implemented as transmons with a dc-SQUID array~\cite{wang2019}, and they are connected with a coupling capacitance.
The frequencies of KPO1 and KPO2 are denoted by $\omega_1$ and $\omega_2$, respectively.
Because of the large detuning, the coupling between the two KPOs is effectively turned off during idle time. To perform the $R_{zz}$ gate, a parametric drive (flux modulation of the dc-SQUID array) for KPO1 with the sum frequency $\omega_3~(=\omega_1+\omega_2)$ is added into the parametric drive with frequency of $2\omega_1$, as shown in Fig.~\ref{fig:2KPO}.
Then, the two-mode squeezing between KPO1 and KPO2 is turned on, because of the three-wave mixing (Fig.~\ref{fig:sum}) due to the nonlinearity of Josephson junctions in KPO1. The two-mode squeezing is known to be sufficient for realizing the $R_{zz}$ gate~\cite{xu2022}. This process is also explained qualitatively as follows. The three-wave mixing leads to difference-frequency generation of a photon of $\omega_2=(\omega_3-\omega_1)$ from a parametric pump photon of $\omega_3$ and a KPO1 photon of $\omega_1$, as shown in Fig.~\ref{fig:sum}. 
The photons of $\omega_2$ propagate to KPO2 via the coupling capacitance.
Since the phase of the generated photons of $\omega_2$ depends on the phase of KPO1, the phase of KPO2 is rotated depending on the phase of KPO1, resulting in the $R_{zz}$ gate ($ZZ$ rotation)~\cite{goto2016_uni,puri2017,goto2019}. 
(A more theoretical explanation is given in Sec.~\ref{sec:two KPO}.)
We thus refer to this gate as a conditional-driving gate.
We also investigate the $R_{zz}$ gate with a difference-frequency drive. (See Appendix B.)

\subsection{
Single KPO}
Before discussing two KPOs, we explain the single-KPO case for simplicity.
We also derive a simple model under some approximations, which is useful for parameter setting.

The SC of a KPO in Fig.~\ref{fig:2KPO} with a time-dependent external magnetic field proportional to $\theta_0-\delta_p\cos\omega_pt$ in the dc-SQUID array is governed by the following Hamiltonian~\cite{wang2019} (See Appendix~\ref{app1}):
\begin{equation}
\label{eq:hami_1KPO}
\hat{H}
=4E_C\hat{n}^2-NE_J\cos(\theta_0-\delta_p\cos\omega_pt)\cos\frac{\hat{\varphi}}{N},
\end{equation}
where $\omega_p$ is the parametric pump frequency ($\omega_p=2\omega_1$ in Fig.~\ref{fig:2KPO}),
$N$ is the number of dc~SQUIDs in the array,
$E_C$ is the charging energy of shunt capacitance, $E_J$ is the Josephson energy,
$\hat{n}$ and  $\hat{\varphi}$ are the Cooper-pair number and phase-difference operators, respectively, satisfying the commutation relation $[\hat{\varphi}, \hat{n}]=\im$.
Equation~(\ref{eq:hami_1KPO}) can be rewritten using bosonic operators $\hat{a}^{\dagger}$ and $\hat{a}$ as 
\begin{equation}
\begin{split}
\label{eq:1KPO-full}    
\hat{H}
&=\omega_1\hat{a}^\dagger\hat{a}
-NE_{J}\cos(\theta_0-\delta_p\cos\omega_{p}t)\cos\frac{\hat{\varphi}}{N} \\
&\quad-NE_{J}\cos\theta_0\cdot\frac{\hat{\varphi}^2}{2N^2},
\end{split}
\end{equation}
where $\omega_1$, $\hat{n}$ and $\hat{\varphi}$ are defined as 
\begin{align}
\label{eq:omega}
\omega_1
&=\left(\frac{8E_C\tilde{E}_{J}}{N}\right)^\frac{1}{2}, \\
\hat{n}
&=\im\left(\frac{\tilde{E}_{J}}{32NE_{C}}\right)^\frac{1}{4}
(\hat{a}^\dagger-\hat{a}), \nonumber\\ \nonumber
\quad\hat{\varphi}
&=\left(\frac{2NE_{C}}{\tilde{E}_{J}}\right)^\frac{1}{4}(\hat{a}^\dagger+\hat{a}),
\nonumber
\end{align}
where $\tilde{E}_{J}~(=E_J\cos\theta_0)$ is the effective Josephson energy.

Assuming a sufficiently small pump amplitude ($\delta_p\ll1$) and performing the rotating-wave approximation (RWA) in a rotating frame of half the pump frequency $\omega_p/2$ and in the transmon limit ($E_C/E_J\ll1$), 
we obtain the Hamiltonian in the simple model as
\begin{equation}
\label{eq:1KPO}
\hat{H}
=\Delta\hat{a}^\dagger\hat{a}
-\frac{K}{2}\hat{a}^{\dagger2}\hat{a}^2
+\frac{P}{2}(\hat{a}^{\dagger2}+\hat{a}^2),
\end{equation}
where $\Delta~(\simeq\omega_1-\omega_p/2)$ is the detuning, and 
the Kerr coefficient $K$ and the parametric pump amplitude $P$ are defined by
\begin{equation}
\label{eq:KP}
K=\frac{E_C}{N^2}, 
\quad P=\delta_p\left(\frac{E_C\tilde{E}_{J}}{2N}\right)^{\frac{1}{2}}\tan\theta_0.
\end{equation} 
With the small amplitude the both models [Eqs.~(\ref{eq:hami_1KPO}) and (\ref{eq:1KPO})] give approximately the same results as shown below.
In this work, we set $\theta_0=\pi/4$ and $\Delta=0$.
The Hamiltonian in Eq.~(\ref{eq:1KPO}) has two degenerate eigenstates $\ket{\pm\alpha}$
with $\alpha=\sqrt{P/K}$~\cite{puri2017,goto2019}, which are used as computational basis states.

\begin{figure*}[t]
\includegraphics[width=160mm]{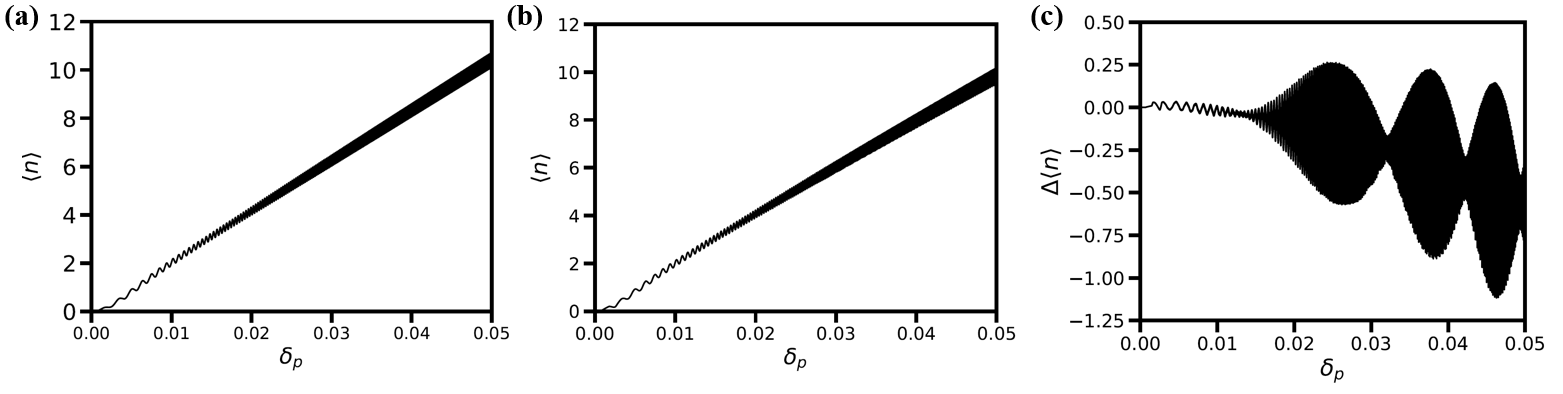}
\caption{Average photon number of a single KPO.
(a) Simple model.
(b) SC model.
(c) Difference between (a) and (b).
The parameters are set as ($E_C/h,~\delta_{p,max},~T,~\omega_1/2\pi,~N$)=(300~MHz, 0.050, 2~$\mu$s, 10~GHz, 5). 
$h$ is the Planck constant. 
The other parameters are obtained by their definitions such as Eqs.~(\ref{eq:omega}) and (\ref{eq:KP}).
}
\label{fig:photon}
\end{figure*}

We calculate the difference in average photon number between the SC and simple models for a single KPO.
In Eq.~(\ref{eq:1KPO-full}), we set the pump frequency as $\omega_p=2\tilde{\omega}_1$, where $\tilde{\omega}_1$ is the single-photon resonance frequency of KPO1 obtained by numerically diagonalizing $\hat{H}$ in Eq.~(\ref{eq:1KPO}). 
We also set the parameters as $E_C/h=300$~MHz, $P=4K$, $\omega_1/2\pi=10$~GHz, and $N=5$.
The other parameters are obtained by their definitions such as Eqs.~(\ref{eq:omega}) and (\ref{eq:KP}).

Figure~\ref{fig:photon} shows the dependence of the average photon numbers in the two models, 
where we solve the Schr\"{o}dinger equation with $\delta_p(t)=\delta_{p,max}t/T$, where $T$ is a sweeping time of $\delta_p$.
Figure~\ref{fig:photon}(c) illustrates the difference between the two models.
The photon number can be increased up to about 10 with a small amplitude such as $\delta_p=0.05$ in both models.
In this region, $\delta_p$ is small enough to suppress the effect of the counter-rotating term, which was investigated in~\cite{masuda2021}.
The fast oscillations with small amplitudes in both models are due to nonadiabatic evolution.
An even smaller oscillation is also shown in Fig.~\ref{fig:photon}(c), which is probably due to fast oscillating terms dropped by the RWA.

\subsection{\label{sec:two KPO}
Two coupled KPOs}
The SC model of the two coupled KPOs shown in Fig.~\ref{fig:2KPO} is given by (see Appendix~\ref{app1} for detail)
\begin{widetext}
\begin{equation}
\begin{split}
\label{eq:2KPO2}    
\hat{H}
&=\sum_{j=1,2}\hat{H}_j+\hat{V}, \\
\hat{H}_1
&=\omega_1\hat{a}_1^\dagger\hat{a}_1-NE_{J1}\cos(\theta_0-\delta_{1}\cos\omega_{p1}t-\delta_g(t)\cos\omega_3t)\cos\frac{\hat{\varphi}_1}{N}-N\tilde{E}_{J1}\cdot\frac{\hat{\varphi}_1^2}{2N^2}, \\
\hat{H}_2
&=\omega_2\hat{a}_2^\dagger\hat{a}_2 
-NE_{J2}\cos(\theta_0-\delta_2\cos\omega_{p2}t)\cos\frac{\hat{\varphi}_2}{N}
-N\tilde{E}_{J2}\cdot\frac{\hat{\varphi}_2^2}{2N^2}, \\
\hat{V}
&=\frac{16E_{C1}E_{C2}}{E_{C0}+E_{C1}+E_{C2}}\hat{n}_1\hat{n}_2,
\end{split}
\end{equation}
\end{widetext}
where $\hat{H}_j$ is the Hamiltonian of KPO$j~(j=1,2)$, $\delta_j$ and $\omega_{pj}$ are the parametric pump amplitude and frequency, respectively, for KPO$j$, 
$E_{C0}$ is the charging energy for the coupling capacitor,
$\delta_g$ is the amplitude of the gate pulse defined as
\begin{equation}
\begin{split}    
\delta_g(t)
=p_{g0}\left(\frac{2N}{E_{C1}\tilde{E}_{J1}}\right)^\frac{1}{2} 
\frac{\tanh\tfrac{\beta t}{T_g}\tanh[\beta(1-\tfrac{t}{T_g})]}
    {\tanh^2\tfrac{\beta}{2}\tan\theta_0},
\end{split}    
\end{equation}
where $p_{g0}$ and $T_g$ denote the peak value of the gate pulse and the gate time, respectively, and $\beta$ is a parameter determining the rise time of the gate pulse.

The Hamiltonian of the simple model is derived in a similar manner to the single-KPO case (See Appendix~\ref{app1} for details):
\begin{equation}
\begin{split}
\label{eq:2KPO}
\hat{H} 
&=\sum_{j=1,2}\hat{H}_j+\hat{H}_\textrm{I}+\hat{H}_g, \\
\frac{\hat{H}_j}{\hbar} 
&=-\frac{K}{2}\hat{a}^{\dagger 2}_j\hat{a}^2_j+\frac{P}{2}
\left(\hat{a}^{\dagger 2}_j+\hat{a}^2_j\right), \\
\frac{\hat{H}_\textrm{I}}{\hbar} 
&=g\left(\hat{a}_1\hat{a}^\dagger_2\e^{-\im\Delta_{12}t}+\hat{a}^\dagger_1\hat{a}_2\e^{\im\Delta_{12}t}\right),  \\
\frac{\hat{H}_g}{\hbar} 
&=\frac{p_g(t)}{2}
\left(\hat{a}^2_1\e^{-\im\Delta_{12}t}+\hat{a}^{\dagger 2}_1\e^{\im\Delta_{12}t}\right),
\end{split}
\end{equation}
where $g$ is a coupling constant (see Appendix~\ref{app1} for its definition), $\Delta_{12}=\omega_1-\omega_2$, $K$ and $P$ are given in the same manner as Eq.~(\ref{eq:KP}), and the gate pulse is given by
\begin{equation}
\label{eq:gate}
p_{g}(t)=p_{g0}\frac{\tanh\tfrac{\beta t}{T_g}\tanh[\beta(1-\tfrac{t}{T_g})]}{\tanh^2\tfrac{\beta}{2}}.
\end{equation}

When $\Delta_{12}$ is sufficiently larger than any other parameters, the effect of $\hat{H}_\textrm{I}$ can be regarded as zero, since the factor $\e^{-\im\Delta_{12}t}$ oscillates very fast.
By the gate pulse described by $\hat{H}_g$, the parametric oscillation amplitudes of KPO1 become $\pm\alpha_1=\pm\sqrt{[P+p_g(t)\e^{\im\Delta_{12}t}]/K}$.
The second term with fast phase rotation corresponds to photons of $\omega_2$ generated by the difference frequency generation (three-wave mixing) in KPO1.
This term cancels out the phase factor in $\hat{H}_\textrm{I}$, resulting in turning on the coupling between the two KPOs. 
This is a mathematical explanation of the conditional driving. Thus, the $R_{zz}$ gate can be performed by the conditional driving induced by the gate pulse.

\section{\label{sec:Rzz}
$R_{zz}$ Gate simulations }
In this section, we show the simulation results of the $R_{zz}$ gate using both the simple and SC models. 

\subsection{
Simple model}
Using the Hamiltonian in Eq.~(\ref{eq:2KPO}), we solve the Schr\"{o}dinger equation to calculate time evolution of the two-KPO state $\ket{\psi}$, and to evaluate the gate fidelity.
For convenience, we introduce the following two state vectors:
\begin{equation}
\begin{split}
&\Ket{\psi_\textrm{even}}\propto(\Ket{\alpha}\otimes\Ket{\alpha}+\Ket{-\alpha}\otimes\Ket{-\alpha}), \\
&\Ket{\psi_\textrm{odd}}\propto(\Ket{\alpha}\otimes\Ket{-\alpha}+\Ket{-\alpha}\otimes\Ket{\alpha}),
\end{split}
\end{equation}
where $\ket{\pm\alpha}$ are coherent states with $\alpha=\sqrt{P/K}$.
We set the initial and ideal final state as 
\begin{align}
\label{eq:psi0}
\Ket{\psi(0)}
&=\mathcal{N}_0(\Ket{\psi_\textrm{even}}+\Ket{\psi_\textrm{odd}}), \\
\label{eq:psi_ideal}
\Ket{\psi_\textrm{ideal}(\Theta)}
&=\mathcal{N}_1(\Ket{\Psi_\textrm{even}}+\e^{\im\Theta}\Ket{\Psi_\textrm{odd}}),
\end{align}
where $\mathcal{N}_0$ and $\mathcal{N}_1$ are normalization factors, and
\begin{equation}
\Ket{\Psi_\textrm{even}}=\hat{U}_0(T_g)\Ket{\psi_\textrm{even}},\quad
\Ket{\Psi_\textrm{odd}}=\hat{U}_0(T_g)\Ket{\psi_\textrm{odd}}.
\end{equation}
Here we have introduced another rotating frame with the time-evolution operator $\hat{U}_0(t)$ without the gate pulse ($\delta_g=0$)~\footnote{This is a technique to extract the effect of the gate pulse, that is, to evaluate the fidelity between the actual state with the gate pulse and an ideal state with an ideal gate operation, where $\hat{U}_0$ (time-evolution operator without the gate pulse) plays a role of the effective identity operation.}.
From the definition of the $R_{zz}$ gate, the final state may be approximately given by
\begin{equation}
\begin{split}
&\Ket{\psi(T_g)}\simeq\alpha_1\Ket{\Psi_\textrm{even}}+\alpha_2\Ket{\Psi_\textrm{odd}},\\
&\alpha_1=\Braket{\Psi_\textrm{even}|\psi(T_g)}, 
 \quad
 \alpha_2=\Braket{\Psi_\textrm{odd}|\psi(T_g)}.
\end{split}
\end{equation}
Hence we define the rotation angle of the $R_{zz}$ gate as $\Theta=\theta_2-\theta_1$, where $\theta_1$ and $\theta_2$ are the arguments of $\alpha_1$ and $\alpha_2$, respectively. 
The gate fidelity of the $R_{zz}$ gate is thus defined by
\begin{equation}
\label{eq:F}
F=|\Braket{\psi_\textrm{ideal}(\Theta)|\psi(T_g)}|^2.
\end{equation}

\begin{figure}[t]
\centering
\includegraphics[height=10cm]
{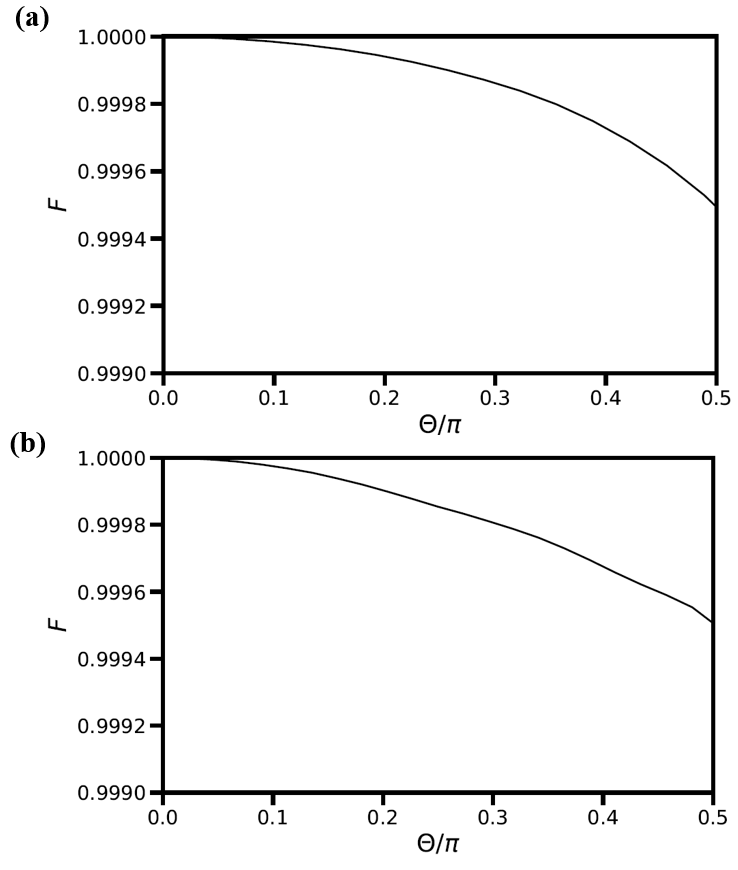}
\caption{Simulation results for $R_{zz}$ gate. 
$F$ and $\Theta$ denote the gate fidelity and the rotational angle, respectively. 
(a) Simple model. 
(b) SC model.
The parameters are set as ($E_{Cj}/h,~P,~\omega_1/2\pi,~\Delta_{12}/2\pi,~g/2\pi,~N,~\beta,~T_g$)=(300~MHz, 4$K$, 10~GHz, $-1$~GHz, 10~MHz, 5, 3, 40~ns). 
The other parameters are obtained by their definitions such as Eqs.~(\ref{eq:omega}) and (\ref{eq:KP}). 
The gate pulse peak $p_{g0}$ is increased up to $5K$.}

\label{fig:F-sum}   
\end{figure}

Figure~\ref{fig:F-sum}(a) shows the simulation results of the gate fidelity using the simple model. 
We find that high fidelities over $99.9\%$ can be achieved for $\Theta$ in the range from $0$ to $\pi/2$, which is sufficient for universality. 
(We have also investigated the $R_{zz}$ gate using a difference-frequency drive~\cite{darwaman2021}, instead of the sum-frequency one, and also obtained high fidelities over $99.9\%$. Interestingly, the fidelity is a little lower than in the case of the sum-frequency drive. See Appendix~\ref{sec:diff} for details.)

\subsection{
SC model}
Next, we perform numerical simulations using the SC model.
Using the Hamiltonian in Eq.~(\ref{eq:2KPO2}), we solve the Schr\"{o}dinger equation and evaluate the gate fidelity, where the initial state is set as Eq.~(\ref{eq:psi0}). 
Using the ideal final state defined by Eq.~(\ref{eq:psi_ideal}) and the final state obtained by the simulation, the gate fidelity is also calculated by Eq.~(\ref{eq:F}).
Figure~\ref{fig:F-sum}(b) shows the gate fidelity of the $R_{zz}$ gate in the SC model.
The $R_{zz}$ gate can be performed with high fidelities over $99.9\%$ in the range from 0 to $\pi/2$.
[The gate fidelity using a SC model with the difference-frequency drive is shown in 
Fig.~\ref{fig:F-diff}(b).]

\section{\label{sec:damping}
Effect of single-photon loss}
\begin{figure}[t]
\centering
\includegraphics[width=8cm]
{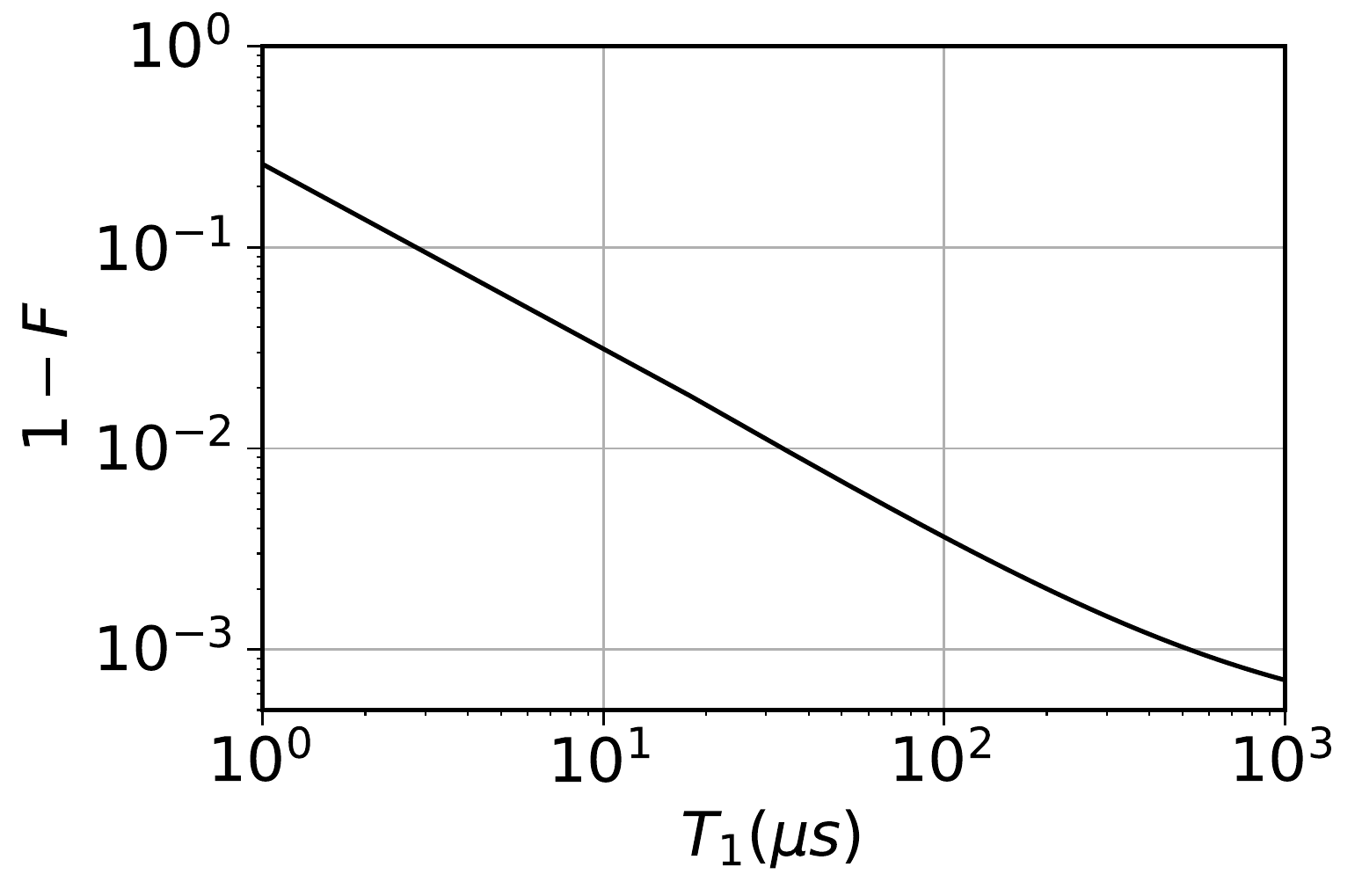}
\caption{Infidelity of $R_{zz}$ gate for the rotation angle $\pi/2$ as a function of single-photon loss rate $T_1=\gamma^{-1}$ using the simple model.
The parameters are the same as in Sec.~\ref{sec:Rzz} A.
}
\label{fig:gamma}   
\end{figure}

Finally, we study the effect of single-photon loss of KPOs, using the simple model. Taking the effect into account, the time evolution of the density operator $\hat{\rho}$ can be written by the following master equation:
\begin{equation}
\partial_t\hat{\rho}
=-\frac{\im}{\hbar}[\hat{H}, \hat{\rho}]+\frac{\gamma}{2}\sum_{j=1,2}
\left(2\hat{a}_j\hat{\rho}\hat{a}_j^\dagger
-\hat{\rho}\hat{a}_j^\dagger\hat{a}_j-\hat{a}_j^\dagger\hat{a}_j\hat{\rho}
\right),
\end{equation}
where $\gamma$ is the single-photon loss rate ($T_1^{-1}$) of KPOs. 
We evaluate the error rates of the $R_{zz}$ gate by solving the master equation with decay time $T_1=\gamma^{-1}$ from 1~$\mu$s to 1~ms. 
Figure~\ref{fig:gamma} shows the infidelity $1-F$ for the rotation angle $\Theta$ of $\pi/2$ as a function of $T_1$, where the fidelity is calculated by 
\begin{equation}
F = 
\braket{\psi_\textrm{ideal}(\Theta)|\hat{\rho}(T_g)|\psi_\textrm{ideal}(\Theta)}.    
\end{equation}
The decay time in a recent experiment~\cite{grimm2020} is $T_1=15.5$~$\mu$s, which corresponds to $1-F\sim2\%$ in Fig.~\ref{fig:gamma}.
Figure~\ref{fig:gamma} also shows that the infidelity below 0.1$\%$ will require a decay time of longer than 500~$\mu$s.
The error due to the single-photon loss can be decreased by shorting the gate time. On the other hand, the shorter the gate time is, the worse the gate fidelity becomes owing to nonadiabatic evolution.
These two factors are in a trade-off relationship on the gate operation. An analysis of this relationship will be carried out in future work.

\section{\label{sec:conc}
CONCLUSION}
In this paper, we have proposed the conditional-driving gate, which is an $R_{zz}$ gate using conditional driving for highly detuned KPOs.

We have also proposed a SC composed of two transmons with a dc-SQUID array (KPOs) coupled with a capacitor.
Using the simple and SC models of the proposed SC, we have numerically demonstrated that the $R_{zz}$ gate can be performed with high fidelity ($>99.9\%$) for rotation angles required for universal quantum computing. 
We have also examined the effect of single-photon loss.
The result suggests that the error probability below 0.1\% will require a decay time of longer than 500 $\mu$s, which seems challenging but is expected to be realized in the future, suggested by recent experimental advances~\cite{wang2022}.
We expect that this simple scheme will be helpful toward the first realization of the two-qubit gate for KPOs.

\section*{Acknowledgements}
We thank Y. Matsuzaki, S. Masuda, T. Ishikawa, and S. Kawabata
for fruitful discussion. This paper is based on results
obtained from a project, JPNP16007, commissioned by
the New Energy and Industrial Technology Development
Organization (NEDO), Japan.

\appendix
\section{\label{app1} 
SC HAMILTONIAN OF TWO KPOs}
The SC Hamiltonian of coupled two KPOs (see Fig.~\ref{fig:2KPO}) in Eq.~(\ref{eq:2KPO2}) is derived from the following Lagrangian:
\begin{equation}
\begin{split}
\mathcal{L}
&=T-U, \\ 
T
&=\sum_{j=1,2}\frac{C_j}{2}\dot{\phi}_j^2+\frac{C_0}{2}(\dot{\phi}_1-\dot{\phi}_2)^2,\quad\\
&=\frac{1}{2}\bm{\dot{\phi}}^tM\bm{\dot{\phi}}, \\
U
&=-\sum_{j=1,2}NE_{Jj}\cos(\theta_0-\delta_j\cos\omega_{pj}t)\cos\frac{\varphi_j}{N},   
\end{split}
\end{equation}
where $T$ and $U$ are the kinetic and potential energy, 
$C_0$ is the coupling capacitance, $C_j$ is the shunt capacitance, $\phi_j$ is the magnetic flux across the dc-SQUID array,
$N$ is the number of dc~SQUIDs in the array, $E_{Jj}$ is the Josephson energy for a single dc~SQUID for KPO$j$.
Here the flux in the dc-SQUID array is modulated with the external magnetic field $\Phi_\textrm{ex}$ such as $\pi\Phi_\textrm{ex}/\Phi_0=\theta_0-\delta_p\cos\omega_pt$ ($\Phi_0$ is the flux quantum). 
The kinetic energy can be rewritten in a compact form using the following flux vector and capacitance matrix:
\begin{equation}
\bm{\dot{\phi}}^t=(\dot{\phi}_1, \dot{\phi}_2),\quad
M=
\begin{pmatrix}
C_1+C_0 & -C_0 \\
-C_0 & C_2+C_0 
\end{pmatrix}.
\end{equation}
The charge $Q_j$, which is the canonical conjugate of $\phi_j$, is given by $\bm{Q}=M\bm{\dot{\phi}},~[\bm{Q}^t=(Q_1, Q_2)]$.
From the Legendre transformation, we obtain the following Hamiltonian:
\begin{equation}
\begin{split}
\label{eq:hami}
H
&=\frac{1}{2}\bm{Q}^tM^{-1}\bm{Q} \\
&\quad-\sum_{j=1,2}NE_{Jj}\cos(\theta_0-\delta_j\cos\omega_{pj}t)
\cos\frac{\varphi_j}{N}.
\end{split}
\end{equation}
Imposing quantization on the Hamiltonian, we obtain the quantum-mechanical Hamiltonian:
\begin{equation}
\begin{split}
\label{eq:H-full-2KPO2}
\hat{H}
&=\hat{H}_0+\hat{V} \\
\hat{H}_0
&=\frac{4E_{C1}(E_{C0}+E_{C2})\hat{n}_1^2+4E_{C2}(E_{C0}+E_{C1})\hat{n}_2^2}{E_{C0}+E_{C1}+E_{C2}} \\
&\quad-\sum_{j=1,2}NE_{Jj}\cos(\theta_0-\delta_j\cos\omega_{pj}t)
\cos\frac{\hat{\varphi}_j}{N}, \\
\hat{V}
&=\frac{16E_{C1}E_{C2}}{E_{C0}+E_{C1}+E_{C2}}\hat{n}_1\hat{n}_2,
\end{split}
\end{equation}
where $\hat{n}_j, \hat{\varphi}_j$ are the Cooper-pair number and phase-difference operators satisfying the relation $[\hat{\varphi}_i, \hat{n}_j]=\im\delta_{i,j}$.
$E_{C0}~(E_{Cj})=e^2/2C_0~(e^2/2C_j)$ is the charging energy of the corresponding capacitance.
Substituting the following into Eq.~(\ref{eq:H-full-2KPO2}), 
\begin{equation}
\begin{split}
\label{eq:aad}
\hat{n}_j
&=\im\left(\frac{\tilde{E}_{Jj}}{32NE_{Cj}}\right)^\frac{1}{4}
(\hat{a}_j^\dagger-\hat{a}_j), \\
\quad\hat{\varphi}_j
&=\left(\frac{2NE_{Cj}}{\tilde{E}_{Jj}}\right)^\frac{1}{4}(\hat{a}_j^\dagger+\hat{a}_j),
\end{split}
\end{equation}
we obtain Eq.~(\ref{eq:2KPO2}). 
Performing the transmon approximation of $\cos(\hat{\varphi}_j/N)$, we also obtain
\begin{widetext}
\begin{equation}
\begin{split}
\label{eq:H-full2}
\hat{H}_0 
&\approx
\sum_{j=1,2}\frac{E_{C0}(E_{C1}+E_{C2})}{E_{C0}+E_{C1}+E_{C2}}\left(\frac{\tilde{E}_{Jj}}{2NE_{Cj}}\right)^\frac{1}{2}
(\hat{a}_j^{\dagger}-\hat{a}_j)^2 
+(E_{Jj}-\delta_j\cos\omega_{pj}t-\delta_g(t)\cos\omega_3t)\left(\frac{E_{Cj}}{2NE_{Jj}}\right)^\frac{1}{2}
(\hat{a}_j^{\dagger}+\hat{a}_j)^2 \\
&\quad-\frac{E_{Cj}}{12N^2}(\hat{a}_j^\dagger+\hat{a}_j)^4, \\
\hat{V}
&=-\underbrace{\frac{4E_{C1}E_{C2}}{E_{C0}+E_{C1}+E_{C2}}\left(\frac{\tilde{E}_{J1}\tilde{E}_{J2}}{4N^2E_{C1}E_{C2}}\right)^\frac{1}{4}}_{\displaystyle{g}} 
(\hat{a}_1^\dagger-\hat{a}_1)(\hat{a}_2^\dagger-\hat{a}_2).
\end{split}
\end{equation}
\end{widetext}
Performing the RWA in a frame rotating at half the pump frequencies, we can obtain the Hamiltonian in Eq.~(\ref{eq:2KPO}).

\section{\label{sec:diff} 
$R_{zz}$ GATE USING DIFFERENCE-FREQUENCY DRIVE}
\begin{figure}[t]
\centering
\includegraphics[width=7.5cm]
{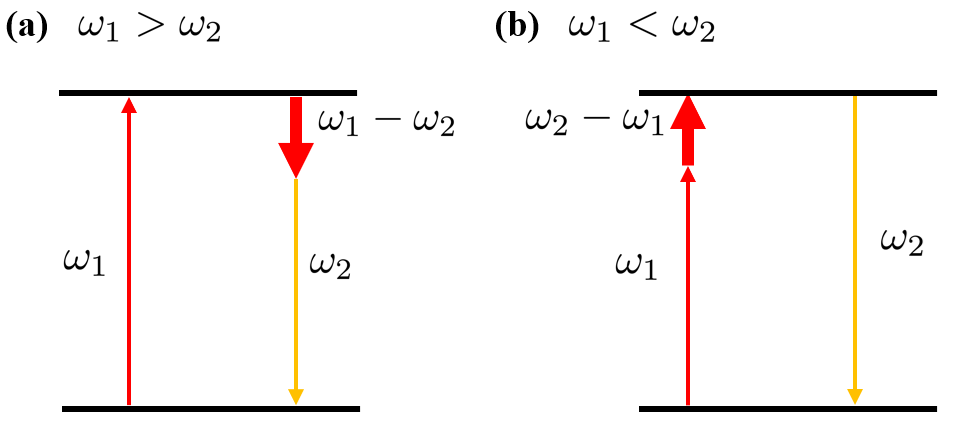}
\caption{Three-wave mixing processes using difference-frequancy drive. 
(a) Difference-frequency generation when $\omega_1 > \omega_2$. 
(b) Sum-frequency generation when $\omega_1 < \omega_2$.
}
\label{fig:diff}
\end{figure}
\begin{figure}[t]
\centering
\includegraphics[height=10cm]
{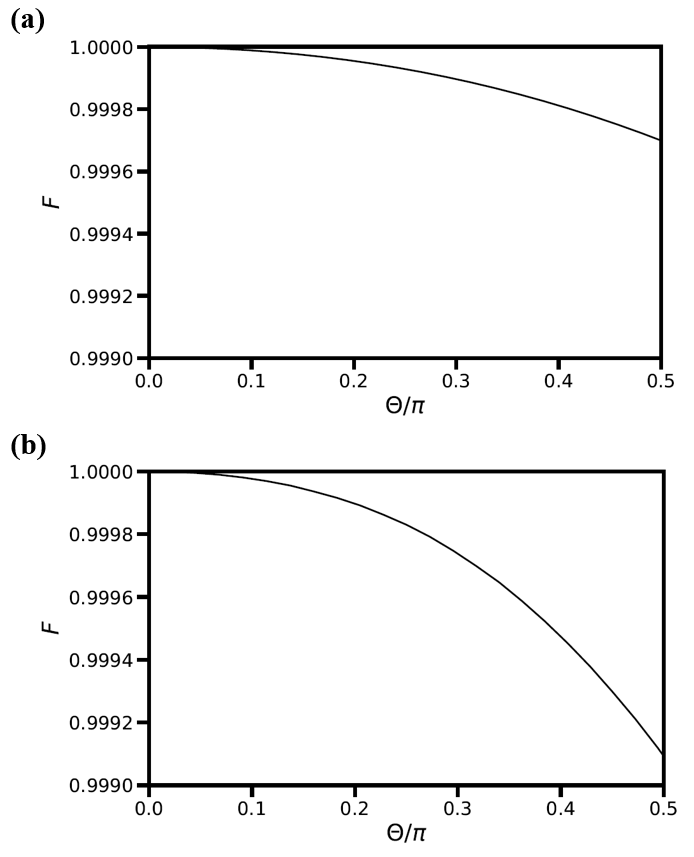}
\caption{Simulation results for $R_{zz}$ gate by the difference-frequency drive when $\omega_1 < \omega_2$ [Fig.~\ref{fig:diff}(b)].  
(a) Simple model. 
(b) SC model. 
$F$ and $\Theta$ denote the gate fidelity and the rotation angle, respectively.
The parameters are set as ($E_{Cj}/h,~P,~\omega_1/2\pi,~\Delta_{12}/2\pi,~g/2\pi,~N,~\beta,~T_g$)=(300~MHz, 4$K$, 10~GHz, $-1$~GHz, 10~MHz, 5, 3, 40~ns).
The other parameters are obtained by their definitions such as Eqs.~(\ref{eq:omega}) and (\ref{eq:KP}). 
The gate pulse peak $p_{g0}$ is increased up to $20K$.
}
\label{fig:F-diff}
\end{figure}

We also investigate the $R_{zz}$ gate using the difference-frequency drive, instead of the sum-frequency drive explained in the main text.
In this case, the three-wave mixing process generating photons of $\omega_2$ differs depending on whether $\omega_1 > \omega_2$ or $\omega_2 > \omega_1$. 
The generation processes are shown in Fig.~\ref{fig:diff}. 
As shown in Fig.~\ref{fig:diff}(a), when $\omega_1>\omega_2$, photons of $\omega_2$ are generated by difference-frequency generation with the gate pulse of $\omega_1-\omega_2$ and photons of $\omega_1$ in KPO1.
On the other hand, when $\omega_2>\omega_1$, the photons of $\omega_2$ are generated by sum-frequency generation with them in Fig.~\ref{fig:diff}(b).

In the case of the difference-frequancy drive, the Hamiltonian in the SC model is also given by Eq.~(\ref{eq:2KPO2}), but in the simple model, $\hat{H}_g$ in Eq.~(\ref{eq:2KPO}) must be replaced by
\begin{equation}
\begin{split}
\hat{H}_g
=\frac{p_g(t)}{2}\cos(\Delta_{12}t)\hat{a}^\dagger_1\hat{a}_1. 
\end{split}
\end{equation}
This Hamiltonian describes both difference-frequency and sum-frequency generations 
shown in Figs.~\ref{fig:diff}(a) and \ref{fig:diff}(b), respectively, because $\cos(\Delta_{12}t)$ has both $\e^{\im\Delta_{12}t}$ and $\e^{-\im\Delta_{12}t}$.

The simulation results of the $R_{zz}$ gate in the simple and SC models are shown in Fig.~\ref{fig:F-diff}.
It is found that we can also achieve high fidelities over $99.9\%$ using the difference-frequency drive.

\newpage
\bibliography{2qubit-KPOs}

\end{document}